# Upper-Lower Bounded-Complexity QRD-M for Spatial Multiplexing MIMO-OFDM Systems

Manar Mohaisen and KyungHi Chang

*The Graduate School of IT and Telecommunications, Inha University,*

*253 Yonghyun-Dong, Nam-Gu, Incheon 402-751, Korea*

Tel: +82 32 860 8422,

Fax: +82 32 865 0480

lemanar@hotmail.com, khchang@inha.ac.kr[1]

Abstract: Multiple-input multiple-output (MIMO) technology applied with orthogonal frequency division multiplexing (OFDM) is considered as the ultimate solution to increase channel capacity without any additional spectral resources. At the receiver side, the challenge resides in designing low complexity detection algorithms capable of separating independent streams sent simultaneously from different antennas. In this paper, we introduce an upper-lower bounded-complexity QRD-M algorithm (ULBC QRD-M). In the proposed algorithm we solve the problem of high extreme complexity of the conventional sphere decoding by fixing the upper bound complexity to that of the conventional QRD-M. On the other hand, ULBC QRD-M intelligently cancels all unnecessary hypotheses to achieve very low computational requirements. Analyses and simulation results show that the proposed algorithm achieves the performance of conventional QRD-M with only 26% of the required computations.

*Keywords: MIMO-OFDM detection, sphere decoding, QRD-M, ULBC QRD-M, maximum likelihood, receiver complexity*

## 1 Introduction

Spatial multiplexing multiple-input multiple-output (MIMO) is an attractive technology due to its capability to linearly increase the system throughput without the need to additional spectral resources [1]. However, the capacity of spatial multiplexing MIMO systems depends mainly on the detection scheme employed at the receiver side to recover the transmitted block of symbols [2]. Consequently, an increasing interest in developing detection techniques to achieve both high performance and low complexity simultaneously has been arisen. Maximum likelihood (ML) detection has been known as the optimum MIMO receiver where the recovery of modulated digital messages requires joint detection of the entire transmitted vector components [3]. In general, to obtain the ML solution, an exhaustive search is needed which is not practical when the problem size is high. Here, the problem size is defined as the number of transmitted symbols via different antennas when the modulation set size is kept fixed.

---

[1] Corresponding Author: KyungHi Chang
The Graduate School of IT & T, Incheon, KOREA 402-751
Tel.: +82-32-860-8422
Fax: +82-32-865-0480
E-mail: khchang@inha.ac.kr
This work was supported by the Korea Science and Engineering Foundation (KOSEF) grant funded by the Korea government (MOST) (No. R01-2008-000-20333-0).

MIMO detection literature includes a variety of detection schemes that differ in strategy adopted, complexity and thus in achieved performance. Among these schemes, sphere decoding (SD) and QR-decomposition with M-algorithm (QRD-M), sometimes referred to as K-best detector, can be considered as the most prominent MIMO detection candidates developed as alternatives to the complex ML detection.
SD achieves a near-ML performance while offering large complexity reduction compared to ML detection [4]-[5]. In [6], the average complexity of SD, under certain assumptions, is shown by Hassibi and Vikalo to be polynomial in the problem size [6]. The problem of SD, however, still resides in its extreme instantaneous complexity which can be attained when the channel matrix is ill-conditioned, i.e., the condition number of the channel matrix is high [7], or when the instantaneous noise power is high.

For practical applications where mobile stations have limited power and low latency tolerance, the random complexity of detection scheme, with high deviation, becomes inacceptable. As a result, fixed complexity QRD-M algorithm was proposed to achieve a compromise between performance and complexity [8], [9]. In the conventional QRD-M algorithm, the required computation is fixed whatever the channel condition and the noise power are. Thus, in the search to obtain constant complexity of the detection process, we neglect the fact that when the channel matrix is well conditioned less computation can be done to achieve exactly the same performance.

In the adaptive QRD-M algorithm proposed in [10], the estimated noise variance multiplied by a constant $X$ summed with the minimum metric at each stage is used as the detection threshold to cancel out unnecessary branches. If $X$ is large, a near-ML performance is achieved while the complexity is increased. On the other hand, as the constant $X$ decreases, the complexity is decreases with degradation in the performance. The problem of adaptive QRD-M algorithm appears at low and medium signal to noise ratio (SNR), where achieving near-ML performance requires high complexity because high number of candidates lead to close metrics and thus will not be cancelled out. Another algorithm was proposed in [11], where threshold at each detection stage is calculated based on the partial decision feedback. Thus, at each detection stage, the node with the minimum metric is used to obtain the solution for the remaining symbols. The resulting metric is used as the new threshold for the current detection stage. Although the complexity is reduced compared to the adaptive QRD-M proposed in [10], the decision feedback process reduces the detector throughput and requires more computational efforts particularly when high number of transmit antennas is employed. Moreover, partial decision feedback algorithm does not have defined method to stop the detection process when obtained solution at first stages is the only possible solution.

In this paper, we propose an upper-lower bounded-complexity QRD-M (ULBC QRD-M) where the size of the search tree is fixed to that of the conventional QRD-M algorithm while intelligently cancelling all branches whose metrics exceed a pre-defined condition. We set, herein, the pre-defined condition to the square Euclidean distance of the Babai point from the received vector. As a result, in certain cases, when the channel matrix is well-conditioned, the Babai point can be the closest point to the received vector, among lattice points survived in the conventional QRD-M algorithm. Thus, the detection process can be terminated at the first detection stage, and consequently complexity is tremendously reduced. On the other hand, when the Babai point is not the closest point, branches which do not satisfy the pre-defined condition are cancelled out and not extended any more. As a consequence, the number of visited nodes decreases, and the complexity is reduced. What is exciting in the proposed algorithm is that it achieves the same performance of the conventional QRD-M algorithm with considerably reduced detection complexity and achieving the same detector throughput.

The remaining part of the paper is organized as follows. In section 2, we explain briefly the system model including the channel model. In section 3, we introduce the ideas behind the SD and the conventional QRD-M algorithm. Section 4 presents the proposed ULBC QRD-M algorithm with derivation of the upper and lower bound complexities in details. In section 5, we show simulation results, and finally we draw the conclusions in section 6.

## 2 System Model

We consider that MIMO spatial multiplexing is applied with orthogonal frequency division multiplexing (OFDM) system with negligible inter-carrier interference. Then, the system can be modeled as

$$\mathbf{r} = \mathbf{H}\mathbf{x} + \mathbf{v}, \quad (1)$$

where $\mathbf{r} \in \mathbb{C}^{N_r}$ is the received vector, $\mathbf{v} \in \mathbb{C}^{N_r}$ is the Gaussian noise withdrawn from i.i.d. centered wide sense stationary process with variance $\sigma^2$. Furthermore, $\mathbf{H} \in \mathbb{C}^{N_r \times N_t}$ is the full column rank complex channel matrix whose element $h_{i,j}$ is the transfer function between the $i^{th}$ receive antenna and the $j^{th}$ transmit antenna.

Consider that $N_r = N_t = N$, and $N_s = 2N_t$ denotes the dimension of the real space. Then, (1) can be mapped to the real domain $\mathbb{R}^{N_s}$ by the following simple equation [12]

$$\begin{bmatrix} \Re(\mathbf{r}) \\ \Im(\mathbf{r}) \end{bmatrix} = \begin{bmatrix} \Re(\mathbf{H}) & -\Im(\mathbf{H}) \\ \Im(\mathbf{H}) & \Re(\mathbf{H}) \end{bmatrix} \begin{bmatrix} \Re(\mathbf{x}) \\ \Im(\mathbf{x}) \end{bmatrix} + \begin{bmatrix} \Re(\mathbf{v}) \\ \Im(\mathbf{v}) \end{bmatrix}, \quad (2)$$

where $\Re(\mathbf{x})$ and $\Im(\mathbf{x})$ are the real and imaginary parts of $\mathbf{x}$, respectively. For clarity, we write (2) as $\mathbf{r}_r = \mathbf{H}_r \mathbf{x}_r + \mathbf{v}_r$, where $\mathbf{r}_r \in \mathbb{R}^{N_s}$, $\mathbf{H}_r \in \mathbb{R}^{N_s \times N_s}$, $\mathbf{v}_r \in \mathbb{R}^{N_s}$ and $\mathbf{x}_r \in \mathbb{Z}^{N_s}$. Operating on the transmitted vector $\mathbf{x}_r$, $\mathbf{H}_r$ generates a lattice $L_s(\mathbf{H}_r) := \{\mathbf{z} = \mathbf{H}_r \mathbf{x}_r \mid \mathbf{x}_r \in \mathbb{Z}^{N_s}\}$. Consequently, (1) can be seen as the result of perturbing the lattice point $\mathbf{z}$ by the noise $\mathbf{v}_r$.

**Table 1** Power and delay profiles for 3GPP SCM-E sub-urban channel model.

| Path  | 1 | 2     | 3     | 4     | 5     | 6     |
|-------|---|-------|-------|-------|-------|-------|
| Delay | 0 | 0.065 | 0.016 | 0.42  | 1.4   | 2.8   |
| Power | 0 | -6.2  | -2.7  | -10.4 | -16.4 | -22.4 |

Finally, using the QR-decomposition, the real channel matrix $\mathbf{H}_r$ is decomposed into the multiplication of a unitary matrix $\mathbf{Q}$ and an upper triangular matrix $\mathbf{R}$. Multiplying both sides of (2) by the inverse of $\mathbf{Q}$, i.e., the transpose, leads to

$$\mathbf{y} = \mathbf{R}\mathbf{x}_r + \mathbf{n}, \quad (3)$$

where $\mathbf{y} = \mathbf{Q}^T \mathbf{r}_r$ and $\mathbf{n} = \mathbf{Q}^T \mathbf{v}_r$. Equation (3) will be used as the starting point to discuss detection algorithms in this paper including the proposed algorithm.

Throughout this paper, we use the 3GPP spatial channel model-extended (SCM-E) detailed in [13]. Briefly, the channel is composed of 6 main-paths that appear in the delay domain as Dirac delta functions; each main-path is summed up of 20 independent sub-paths where sub-paths are grouped into mid-paths, and each main-path is composed of 3 or 4 mid-paths depending on the channel scenario. In what follows, we consider sub-urban macro scenario in which each main-path contains 3 mid-paths. Table I gives the power and the delay profiles at carrier frequency $f_c$ = 3.7 GHz. Delay profile is given micro seconds, and the power of the paths is given with respect the power of the first path in dB.

## 3 Conventional Detection Algorithms

### 3.1 Sphere Decoding

The two non-trivial strategies on which SD is based are introduced by Fincke and Pohst [14] and Schnorr-Euchner [15]. We restrict our discussions herein on the strategy of [15]. Based on the strategy of the Schnorr-Euchner strategy, SD tests hypotheses which satisfy the condition

$$\left\| \mathbf{R}\mathbf{x}_r - \mathbf{y} \right\|^2 \leq d^2, \quad (4)$$

where $d^2$ is the pre-defined square radius of the search sphere. SD finds the components of the solution vector in a successive way. In other words, the exhaustive search in the Ns-dimensional sphere, i.e., ML, is transformed into $N_s$ 1-dimensional search problems. As a consequence, hypotheses which don't satisfy (4) at any detection stage are cancelled out, and thus the size of the search tree is reduced.

When the channel matrix is well-conditioned and instantaneous noise power is low, SD can achieve a tremendous computational economy while achieving a near-ML performance. The drawback of SD appears when the channel matrix is ill-conditioned or when the instantaneous noise power is high. In this case, many hypotheses result in similar square Euclidian distances from the received vector, and thus SD becomes incapable to cancel any of these hypotheses at early stages of the detection process.

Finally, we can conclude that in spite of the appreciable reduction in the expected complexity by SD, the complexity of the extreme case is still a handicap that prevents the implementation of SD in commercial communication systems due to the requirements of power and delay tolerance.

In the next section, QRD-M with fixed complexity is discussed as a solution to the extreme case of SD complexity.

### 3.2 Conventional QRD-M Algorithm

QRD-M algorithm retains only a pre-defined number of branches, $M_i$, at the $i^{th}$ detection stage instead of retaining all hypotheses satisfying (4). As a result, the complexity of the algorithm becomes deterministic for fixed problem size and $\boldsymbol{M} = [M_{Ns}, M_{Ns-1}, …, M_1]$.

In this paper, we reformulate the QRD-M algorithm in the real-domain as following steps:

1. Start from the root of the search tree with zero accumulative metric, and set $i = N_s$ which is the first detection stage.
2. Extend all retained branches to all possible nodes.
3. Order resulting branches based on their accumulative metrics defined in (5), and retain the best $M_i$ branches with the least branch metrics.
4. If the last detection stage is reached ($i = 1$), go to "step 6", otherwise continue.
5. Move to the next stage ($i = i - 1$), and go to "step 2".
6. Order survival branches based on their accumulative metric and retain the best branch as the solution of QRD-M algorithm.

The accumulative metric of (4) is given in the expanded summation form as following:

$$\sum_{i=1}^{N_s} \left| y_i - R_{i,i}\,\widehat{x}_i - \sum_{j=i+1}^{N_s} R_{i,j} x_j \right|^2, \qquad (5)$$

where $R_{k,j}$ is the element of the matrix **R** located at the $k^{th}$ row and the jth column. Also, $i$ indicates the detection stage at which the metric is calculated, and $x$ is the demodulation of the zero-forcing solution $\widehat{x}$. Note that the accumulative metric at the first detection level, i.e., $i = N_s$, is given as $\left( y_{Ns} - R_{Ns,Ns}\hat{x}_{Ns} \right)^2$.

## 4 Proposed ULBC QRD-M Algorithm

SD achieves the closest performance to the ML detection. Its drawback, however, resides in its worst-case complexity which is comparable to the complexity of ML. QRD-M algorithm represents a compromise between complexity and performance, where the size of the tree search is fixed at the beginning of the detection process. The drawback of QRD-M algorithm is that it retains fixed number of branches at each detection stage regardless of their accumulative metric values. Therefore, when the channel matrix is well-conditioned, unnecessary computations are always being done.

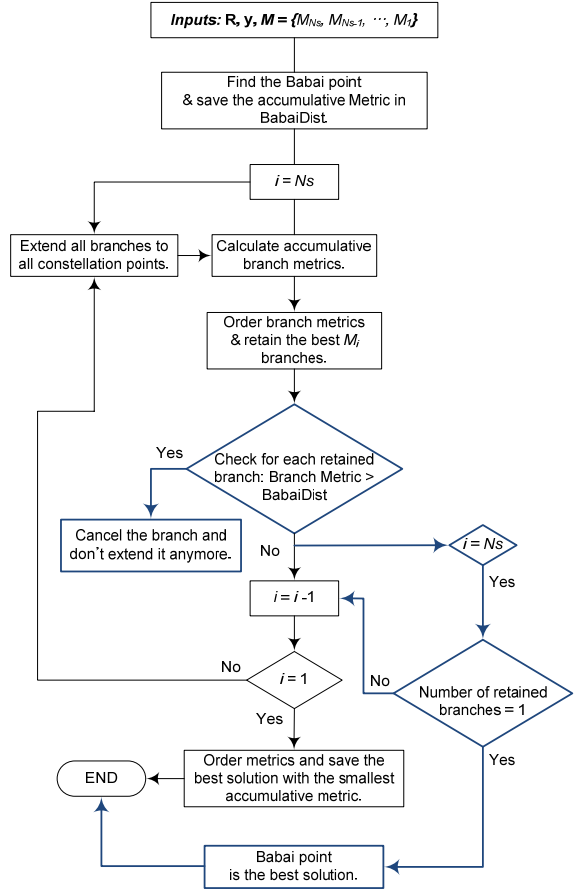

Fig. 1 Proposed ULBC QRD-M algorithm.

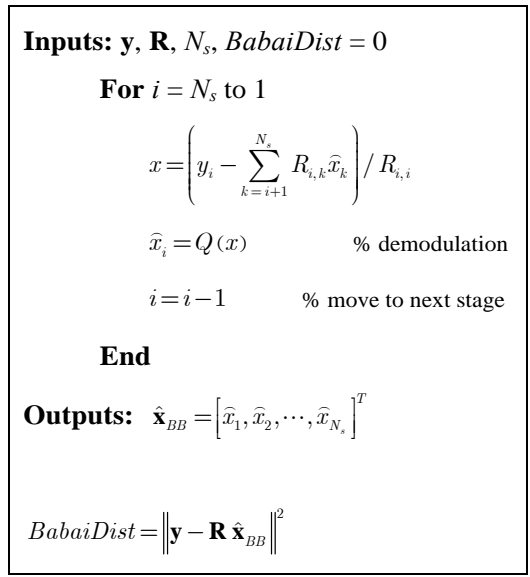

Algorithm 1. Babai point solution.

In this paper, we propose a modified QRD-M algorithm, named as ULBC QRD-M, which fixes the upper computational requirements to that of the conventional QRD-M algorithm, while cancelling all branches which don't satisfy a pre-defined condition. Fig. 1 gives the flowchart of the proposed ULBC QRD-M algorithm, where the inputs to the proposed algorithm are the **R** matrix, **y** and the parameter *M*.

The proposed ULBC QRD-M algorithm is described in the following steps.

1. Find the Babai point solution using the algorithm depicted in Algorithm 1 [16], i.e., simple successive interference cancellation starting by the stage $N_s$, calculate its accumulative square Euclidian distance from **y** using the left part of (4), save it in *BabaiDist*, and then,
    - Extend the root node of the search tree to all possible branches.
    - Calculate branch accumulative metrics and retain the best $M_i$ branches.
    - Compare the accumulative metrics of the retained branches with *BabaiDist*, and cancel branches with higher metrics.
    - If the number of survival branches = 1, detection algorithm is finished, otherwise continue.
2. Move to the next detection stage ($i = i - 1$).
3. Extend all retained branches to all possible nodes.
4. Order resulting branches based on their accumulative metrics, and retain the best $M_i$ branches.
5. Compare the accumulative metrics of the survival branches with *BabaiDist*, and cancel branches with larger accumulative metrics.
6. If the last detection stage is reached (i.e., $i = 1$), order survival branches based on their accumulative metrics, and save the best as the solution of ULBC QRD-M algorithm. Otherwise, go to "step 2".

The choice of the *BabaiDist* as the initial radius is motivated by the following:

1. Calculating the *BabaiDist* returns a solution which is used to finish the detection process in "step 1", if the Babai point is the best among the lattice points retained by the conventional QRD-M.
2. Other criterions to define the search radius [17] can lead to an empty set of solutions, thus the radius must be increased and detection is re-initiated. On the other hand, the search radius can be large in such a way many lattice points are included, and consequently the computational complexity increases.

The computational complexity of the proposed ULBC QRD-M algorithm is explained in the example of Fig. 2.

The general scenario is shown in Fig. 2(a), where the Babai point is closer to the received vector than a set of points returned by the conventional QRD-M algorithm. So, these points are cancelled out during the detection process, and thus the average number of visited nodes is reduced.

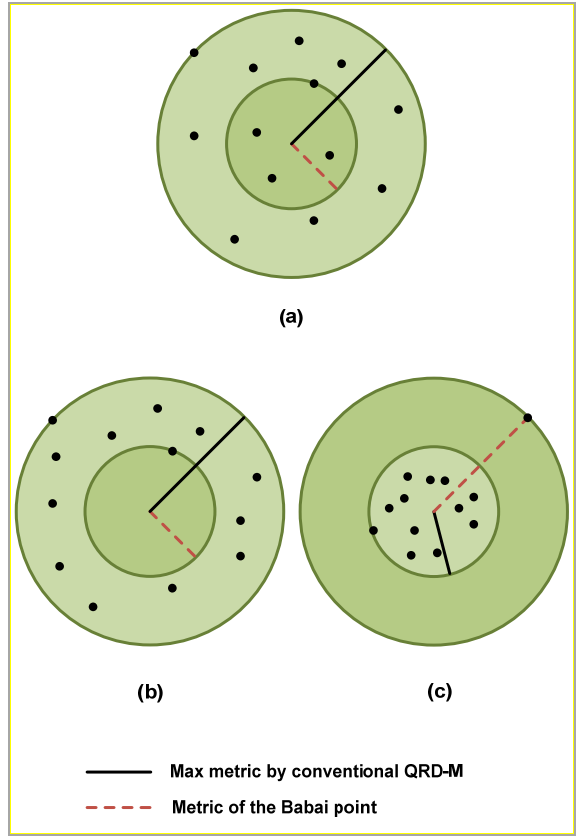

Fig. 2. Complexity of the proposed ULBC QRD-M algorithm: (a) General scenario (b) Best scenario (c) Worst scenario.

The lower bound on the complexity, which is the best case, can be achieved when the Babai point is the closest lattice point to the received vector as shown in Fig. 2(b). Then, the lower bound on the number of visited nodes is given by

$$f_{ULBC}^{LB} = N_s + C^{1/2}, \quad (6)$$

where $C$ is the modulation set size, e.g., 4 for QPSK, and $N_s$ is the number of visited nodes to obtain the Babai point.

The upper bound on the complexity, which is the worst case, is achieved when the Babai point is further from the received vector than the furthest retained point by the conventional QRD-M algorithm as depicted in Fig. 2(c). As a result, the upper bound on the number of visited nodes is given by

$$f_{ULBC}^{UB} = f_{QRD-M} + N_s, \quad (7)$$

where $f_{QRD-M}$ is the number of visited nodes by the conventional QRD-M algorithm. As compared to the conventional QRD-M algorithm, the proposed algorithm maintains the computational complexity of the QRD-M as its upper bound, while it adaptively avoids unnecessary computations that are done in the case of the conventional QRD-M. Note that the performances of both algorithms are the same as shown in Section 5.

# 5 Simulation Results and Discussion

In this section, we verify the performance of the proposed ULBC QRD-M algorithm compared to those of the conventional SD and QRD-M algorithms.

Table 2 Simulation parameters.

| Parameter | Value |
|---|---|
| Bandwidth ($BW$) | 20 MHz |
| Sampling frequency ($f_s$) | 30.72 MHz |
| Carrier frequency ($f_c$) | 3.7 GHz |
| Subcarrier spacing ($\Delta f$) | 15 kHz |
| FFT size ($N_{FFT}$) | 2048 |
| Data sub-carriers ($N_{DC}$) | 1200 |
| Guard interval ($GI$) | 146 samples |
| Sub-frame duration ($T_s$) | 7 OFDM symbols (0.5 ms) |
| Modulation | 16 QAM |
| Channel coding | Turbo code (8 decoding iterations) |
| Code rate | 1/2 |
| Packet length | 2400 bits |
| MIMO channel model | 3GPP spatial channel model-extended (SCM-E) |
| Channel scenario | Suburban macro |
| Mobility | 120 km/h |
| Channel estimation | Pilot-assisted |
| Detection algorithms | SD, QRD-M, Proposed ULBC QRD-M |
| Number of Tx/Rx antennas | 4×4 |

Table 2 lists the main simulation parameters used in this paper. The Turbo coder is a parallel concatenation of two 1⁄2 code rate binary recursive systematic convolutional (RSC) encoders with feedforward and feedback polynomials 0xd and 0xb, respectively.

The RSC encoders are separated by a random interleaver. Furthermore, puncturing is used to increase the code rate of the turbo encoder from 1⁄3 to 1⁄2.

In this paper, the zero-forcing QR-decomposition [18] has been used for all the detection algorithms. Also, the same signal ordering is used to calculate the Babai point since only one QR-decomposition is calculated. Furthermore, the initial radius d in the sphere decoder has been set to infinity since it does not affect the computational complexity of the algorithm when the Schnorr-Euchner enumeration is used (see [15] for more details).

Fig. 3 depicts the transmission frame structure when 4 transmit antennas are used. Pilots of different antennas are orthogonal with each other and inserted in the 2nd and 6th OFDM symbols. The channel is estimated at pilot positions by means of ZF algorithm, and then channel coefficients at data positions are estimated by performing interpolation in both the time and the frequency domains. This frame structure can be used for any number of transmit antennas by manipulating the density of pilot.

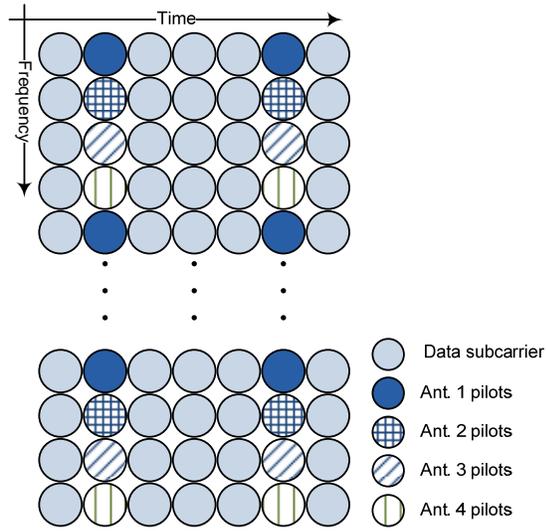

Fig. 3 Frame structure for 4×4 MIMO-OFDM system.

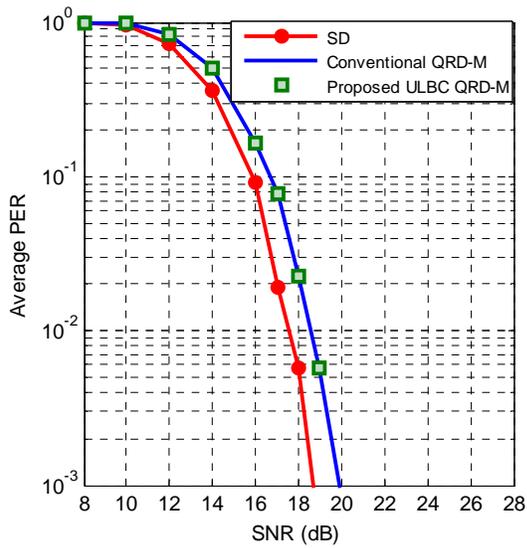

Fig. 4 Performance of the proposed ULBC QRD-M compared to those of conventional algorithms.

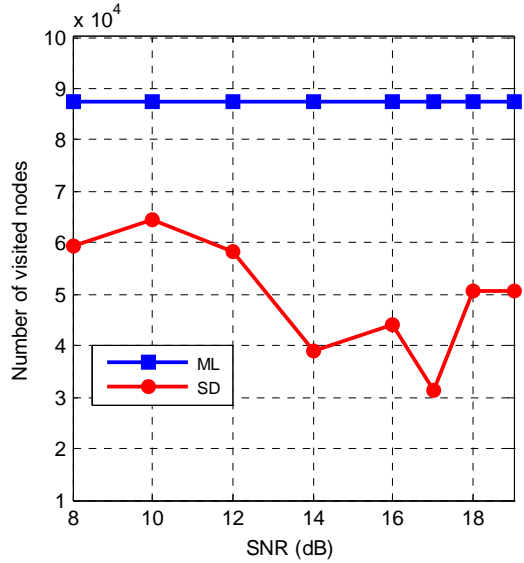

Fig. 5 Worst case complexity of SD algorithm.

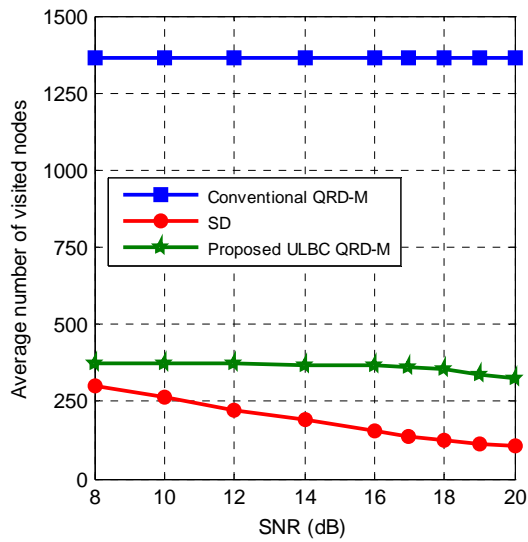

Fig. 6 Average complexity of the proposed ULBC QRD-M algorithm.

Fig. 4 shows the PER performance of the proposed ULBC QRD-M compared to SD and conventional QRD-M algorithms. The proposed ULBC QRD-M algorithm and the conventional QRD-M algorithm show the same PER performance. This implies that the proposed algorithm does not deviate from the solution obtained by conventional QRD-M.

At target PER $10^{-2}$, SD outperforms both the conventional QRD-M and ULBC QRD-M by about 1 dB in signal to noise ratio (SNR), where ***M*** = [$q, q^2, q^3, ..., q^3$] and $q = (C)^{0.5} = 4$ for 16 QAM.

Fig. 5 shows the simulated extreme complexity of SD for 16 QAM and 4×4 MIMO-OFDM system together with the ML detection complexity. The main drawback of SD resides in its extreme complexity which exceeds 60% of the ML complexity when the search sphere includes a high number of solutions or when the channel matrix is ill-conditioned.

Fig. 6 shows the average complexity of the proposed ULBC QRD-M algorithm compared to the average complexities of both SD and conventional QRD-M. SD achieves the best average complexity while conventional QRD-M algorithm has a fixed but respectively high complexity. The proposed algorithm achieves the performance of QRD-M algorithm with only 26% of the expected computational requirements. At SNR = 8 dB, ULBC QRD-M (see Fig. 6), adaptive QRD-M ($X$ = 4.0) [10], and partial decision feedback algorithms (see Fig. 5 in [11]) achieve the same performance of the conventional QRD-M algorithm with 27.6%, 93.15%, and 47.2% of the complexity of QRD-M algorithm. Furthermore, the average complexity of the proposed algorithm has a small variance in the operational range of SNR, compared to both adaptive and decision feedback algorithms of [10] and [11]. This implies that the average latency of the proposed algorithm is almost fixed while those of both the pre-mentioned algorithms are much variant depending on SNR. Moreover, the upper bound on the complexity of the proposed algorithm is lower than those of both adaptive and decision feedback algorithms because of the required calculations for obtaining the thresholds.

Furthermore, the most important benefit of the proposed algorithm is not only its low average complexity but also its upper bounded complexity as defined in (7). The low average complexity of the proposed algorithm is due to cancelling all branches which have accumulative metrics larger than that of the Babai point. This leads, in some cases, to terminate the detection process after calculating few metrics.

# 6 Conclusions

In this paper, we proposed an upper-lower bounded-complexity detection algorithm for spatial multiplexing MIMO-OFDM systems. In the proposed algorithm, the size of the search tree is pre-defined to limit the upper bound complexity of the proposed algorithm. On the other hand, unnecessary computations are omitted by fixing the radius of the search sphere in such a way it includes the best solution. As simulation results show the proposed ULBC QRD-M algorithm achieves exactly the same performance as the conventional QRD-M algorithm while reducing the expected computational complexity by 74%. In light of the presented analyses and obtained simulation results, we conclude that the proposed ULBC QRD-M algorithm can be considered as a prominent replacement for both SD and conventional QRD-M for practical applications.

# References

1. S. Haykin and M. Moher, Modern Wireless Communications. Pearson Prentice Hall, New Jersey, 2005.

2. K. Su, Efficient Maximum Likelihood Detection for Communication over Multiple Input Multiple Output Channels. Doctoral Dissertation, University of Cambridge, 2005.

3. J. Jalden and B. Ottersten, "On the complexity of sphere decoding in digital communications," IEEE Trans. on Sig. Processing, pp. 1474-1484, Apr. 2005.

4. J. Boutros, N. Gresset, L. Brunel, and M. Fossorier, "Soft-input soft-output lattice sphere decoder for linear channels," in Proc. GLOBECOM, 2003, pp. 1583-1587.

5. B. M. Hockwald and S. ten Brink, "Achieving near-capacity on a multiple-antenna channel," IEEE Trans. on Communications, pp. 389-399, Mar. 2003.

6. B. Hassibi and H. Vikalo, "On the expected complexity of integer least-squares problems," in Proc. IEEE ICASSP, May 2002, pp. 1497-1500.

7. G. Golub and C. F. Van Loan, Matrix Computations. The Johns Hopkins University Press, London, 1996.

8. J. Anderson and S. Mohan, "Sequential coding algorithms: A survey and cost analysis," IEEE Trans. on Communications, pp. 169-176, Feb. 1984.

9. K.-J. Kim and R. Iltis, "Joint detection and channel estimation algorithms for QS-CDMA signals over time-varying channels," IEEE Trans. on Communications, pp. 845-855, May 2002.

10. H. Kawai, K. Higuchi, N. Maeda, and M. Sawahashi, "Adaptive control of surviving symbol replica candidates in QRD-MLD for OFDM MIMO multiplexing," IEEE J. Sel. Areas Communications, pp. 1130-1140, June 2006.

11. K. Jeon, H. Kim, and H. Park, "An efficient QRD-M algorithm using partial decision feedback detection," in Proc. 40th Asilomar Conference on Signal, Systems, and Computers, Oct 2006, pp. 1658-1661.

12. H. Vikalo, B. Hassibi, and T. Kailath, "Iterative decoding for MIMO channels via modified sphere decoding," IEEE Trans. on Wireless Communications, pp. 2299-2311, Nov. 2004.

13. D. S. Baum, J. Salo, G. Del Galdo, M. Milojevic, P. Kyösti, and J. Hansen, "An interim channel model for beyond-3G systems," in Proc. IEEE VTC, May 2005.

14. U. Fincke and M. Pohst, "Improved methods for calculating vectors of short length in a lattice, including complexity analysis," Mathematics of Computation, pp. 463-471, Apr. 1985.

15. E. Agrell, T. Eriksson, A. Vardy, and K. Zeger, "Closest point search in lattices," IEEE Trans. on Inf. Theory, pp. 2201-2214, Nov. 2002.

16. L. Babai, "On Lovasz' lattice reduction and the nearest point problem," Combinatorica, pp. 1-13, Mar. 1986.

17. Y. Dai, S. Suni, and Z. Lei, "A comparative study of QRD-M detection and sphere decoding for MIMO-OFDM systems," in Proc. PIMRC, 2005, pp. 186-190.

18. D. Wubben, R. Bohnke, V. Khun, and K.-D. Kammeyer, "Efficient algorithm for decoding layered space-time codes," Electronics Letters, pp. 1348-1350, October 2001.